\documentclass[traditabstract,onecolumn]{aa}
\usepackage{txfonts} 
\usepackage[numbers]{natbib}
\usepackage{graphicx}
\pdfoutput=1

\newcommand{\Msun}{\ensuremath{M_{\sun}}}
\newcommand{\mbh}{\ensuremath{M_{\rm BH}}}
\newcommand{\mnsc}{\ensuremath{M_{\rm NSC}}}
\newcommand{\mstar}{\ensuremath{M_{\star}}}
\newcommand{\mtot}{\ensuremath{M_{\star, \, {\rm tot}}}}
\newcommand{\mbulge}{\ensuremath{M_{\star, \, {\rm bul}}}}

\newcommand{\sigzero}{\ensuremath{\sigma_{0}}}

\newcommand{\hi}{\ion{H}{i}}

\begin{document}

\title{Do Nuclear Star Clusters and Supermassive Black Holes Follow the
Same Host-Galaxy Correlations?\thanks{To appear in \textit{Advances in Astronomy} special issue:
``Seeking for the Leading Actor on the Cosmic Stage: Galaxies versus 
Supermassive Black Holes''}}

\titlerunning{NSCs and SMBHs: Galaxy Correlations}

\author{Peter Erwin\inst{1,2}
\and Dimitri Alexei Gadotti\inst{3}}

\institute{Max-Planck-Institut f\"{u}r extraterrestrische Physik,
Giessenbachstrasse,
D-85748 Garching, Germany
\and Universit\"{a}ts-Sternwarte M\"{u}nchen,
Scheinerstrasse 1,
D-81679 M\"{u}nchen, Germany
\and European Southern Observatory,
Alonso de Cordova 3107, Vitacura,
Casilla 19001, Santiago 19, Chile}

\date{Received <date> / Accepted <date>}

\abstract{
Studies have suggested that there is a strong correlation between the masses of
nuclear star clusters (NSCs) and their host galaxies, a correlation which said
to be an extension of the well-known correlations between supermassive black
holes (SMBHs) and their host galaxies. But careful analysis of disk galaxies --
including 2D bulge/disk/bar decompositions -- shows that while SMBHs correlate
with the stellar mass of the \textit{bulge} component of galaxies, the masses of
NSCs correlate much better with the \textit{total} galaxy stellar mass. In
addition, the mass ratio $\mnsc/\mtot$ for NSCs in spirals (at least those with
Hubble types Sc and later) is typically an order
of magnitude smaller than the mass ratio $\mbh/\mbulge$ of SMBHs.  The absence of
a universal ``central massive object'' correlation argues against common
formation and growth mechanisms for both SMBHs and NSCs. We also discuss
evidence for a break in the NSC--host galaxy correlation: galaxies with Hubble
types earlier than Sbc appear to host systematically more massive NSCs than do
types Sc and later.
}

\maketitle

\section{Introduction}

As far as we can tell, all massive galaxies in the local universe harbor
supermassive black holes (SMBHs, with masses $\mbh \sim 10^{6}$--$10^{9}
\Msun$). The masses of these SMBHs correlate strongly with several global
properties of the host galaxies, particularly with the central velocity
dispersion \sigzero{} \citep{ferrarese00,gebhardt00} and with the bulge luminosity or mass
\citep[e.g.,][]{marconi03,haring04}. These correlations imply that the processes
which drive galaxy growth and the processes which drive black hole growth are
intimately linked -- perhaps even the \textit{same} processes.

It is now also clear that many galaxies, particularly later-type spirals, host
luminous nuclear star clusters \citep[NSCs; e.g.,][]{carollo97,boker02}, with
masses in the range $10^{5}$--$10^{8} \Msun$; see the review by \citet{boker08}
for more details. Recently, several authors have argued that NSCs and central
SMBHs have the \textit{same} host-galaxy correlations: in particular, that SMBHs
and NSCs have the same correlation with bulge luminosity and mass
\citep{wehner06,ferrarese06,cote06,rossa06} (but see \citet{balcells07}).  The
suggestion, then, is that NSCs and SMBHs are in a sense members of the same
family of ``Central Massive Objects'' (CMOs), and thus that they may have grown
via the same mechanisms
\citep[e.g.,][]{mclaughlin06,li07,nayakshin09,devecchi10}.

We argue, however, that one should be cautious about assuming that NSCs and
SMBHs are really part of the same family, with the same host-galaxy
relationships. To begin with, the samples of \citet{wehner06} and
\citet{ferrarese06}, which were used to make the CMO argument, were almost
entirely early-type galaxies -- mostly ellipticals and dwarf ellipticals. These
are galaxies which are, in essence, ``pure bulge'' systems, so one could just as
easily argue for a correlation with \textit{total} galaxy mass.  But we know
that SMBHs in \textit{disk} galaxies correlate better with just the bulge, and
\textit{not} with the total galaxy mass or light
\citep[e.g.,][]{kormendy-gebhardt01,kormendy11}. Given that there have been
previous claims that NSCs in spiral galaxies correlate with the \textit{total}
galaxy light \citep[e.g.,][]{carollo98}, we are prompted the ask the question:
do nuclear clusters \textit{in disk galaxies} correlate with the bulge (like
SMBHs), or with the whole galaxy?


\section{Samples, Methodology, and Data Sources}

Although current studies suggest that the \mbh-\sigzero{} relation is tighter
and has less intrinsic scatter than the \mbh-\mbulge{} relation
\citep[e.g.,][]{gueltekin09b}, velocity dispersion is \textit{not} the ideal
host-galaxy measure to use here, for three reasons. First, most of the
best-determined NSC masses are based directly on the measured velocity
dispersion of the NSC \citep[e.g.,][]{walcher05}, which is often
indistinguishable from that of the surrounding bulge; this means a (spurious)
correlation between NSC mass and central velocity dispersion is only to be
expected. Second, some NSCs are found in galaxies with \textit{no} detectable
bulge at all (see discussion in Section~\ref{sec:comparison}).  Finally, it is
difficult to see how one should discriminate between a velocity dispersion due
to the bulge versus one due to the whole galaxy. But discriminating between
bulge and whole-galaxy luminosities and masses is much simpler. So we choose
instead to compare NSCs and their host galaxies with the \mbh-\mbulge{}
relation, which means comparing NSC masses with the \textit{stellar masses} of
host galaxies and their bulges.

For NSCs, we emphasize galaxies where the NSC masses have been
\textit{dynamically} measured, since this is the most direct analog to
well-determined SMBH masses (i.e., those with direct dynamical mass measurements
from stellar, gas, or maser kinematics, where the SMBH sphere of influence is
resolved).  In addition, dynamical measurements avoid possible
problems with multiple stellar populations; the latter can potentially bias
stellar masses estimated from broad band colors.  Spectroscopic studies
\citep{seth06,rossa06} have shown that NSCs often contain multiple stellar
populations; this renders mass estimates based on Single Stellar Population
(SSP) models \citep[e.g., those used by][]{ferrarese06} somewhat uncertain. The
NSCs we focus on are taken primarily from the sample of \citet{walcher05}, with
additional data from \citet{ho96}, \citet{boker99}, \citet{kormendy-bender99},
\citet{matthews99} and \citet{gebhardt01}, \citet{barth09}, \citet{seth10}, and
\citet{kormendy10}; we use the estimate of \citet{launhardt02} for the Milky
Way's NSC. This gives us a total of 18 galaxies with dynamically determined NSC
masses.  These cover Hubble types S0--Sm, but the sample is in fact heavily
biased towards later types: over three-quarters are Hubble types Scd or later.
As an additional, secondary sample, we include 15 galaxies from \citet{rossa06},
where the masses are estimated by fits of multiple SSP models to high-resolution
spectroscopy. Most of these galaxies are Sc and later, but a few earlier-type
spirals (Sa--Sb) are also included.

Total stellar masses are based on $K$-band total magnitudes from 2MASS
\citep{jarrett00} or from \citet{malhotra96} for M31 and M33 (which are too
large for accurate sky subtraction of 2MASS images), combined with color-based
mass-to-light ($M/L$) ratios from \citet{bell03}. For the latter we use optical
colors from the literature (primarily from
HyperLeda\footnote{http://leda.univ-lyon1.fr}) or from direct measurements on
Sloan Digital Sky Survey \citep[SDSS,][]{york00} images.  The \textit{bulge}
masses are derived using bulge-to-total ($B/T$) values determined individually
for each galaxy by 2D image decomposition, using the BUDDA software package
\citep{desouza04,gadotti08}, which incorporates bulge and disk components
\textit{and} optional bars and central point sources (the latter can be used for
both nuclear star clusters and AGN). Note that we explicitly define ``bulge'' to
be the ``photometric bulge'' -- that is, the excess light (and stellar mass)
which is not part of the disk, bar, or nuclear star cluster. We defer questions
of how SMBH (or nuclear cluster) mass relates to so-called ``pseudobulges''
versus ``classical bulges'' \citep[e.g.,][]{hu08,nowak10} to a later analysis.


Full 2D decompositions, as described above, were used for all S0 and spiral SMBH
host galaxies. For the NSC host galaxies, we follow the same approach, with one
simplification. Since we have found that $B/T$ ratios for \textit{unbarred}
galaxies do not change dramatically if we use 1-D surface-brightness profile
decompositions instead of 2D image decompositions, we use the former for
genuinely unbarred galaxies; we are careful to exclude (or separately model) the
NSC contribution to the surface-brightness profile in these cases. Galaxies
which \textit{do} possess bars are subjected to full 2D decompositions; see the
following section for details.

\subsection{Bulge-Disk Decompositions}

As noted previously, we use 2D image decompositions via the BUDDA software
package to determine the $B/T$ ratios, and thus the bulge stellar masses, for
SMBH host galaxies and for barred NSC host galaxies. For the NSC galaxies, we
use \textit{HST} data wherever possible, to enable the NSC itself to be properly
modeled as a separate source. However, we have found that when the NSC is
sufficiently luminous, and when the bulge is sufficiently low-contrast, we can
achieve reasonable decompositions with ground-based images; these are sometimes
preferable if they are near-IR (to minimize the effects of dust extinction and
recent star formation) and/or large enough to include the entire galaxy (to
allow better recovery of the disk component).

%
%
%
%

We have completed decompositions for the galaxies with dynamically determined
NSC masses (we use the published 2D decomposition of \citep{barth09} for
NGC~3621); in the special case of the Milky Way, we assume a bulge mass of $\sim
1.0 \times 10^{10} \Msun$ and a total stellar mass of $5.5 \times 10^{10}
\Msun$, based on arguments in \citet{dehnen-binney98}, \citet{klypin02}, and
\citet{flynn06}. Decompositions for the spectroscopic sample are still in
progress, but are mostly complete; since our primary analysis (e.g., computing
the $\mnsc$-$\mbulge$ relation) is based on the dynamical masses, the
incompleteness of the spectroscopic sample does not affect our results.

\begin{figure*}
\centering
\includegraphics[width=17cm]{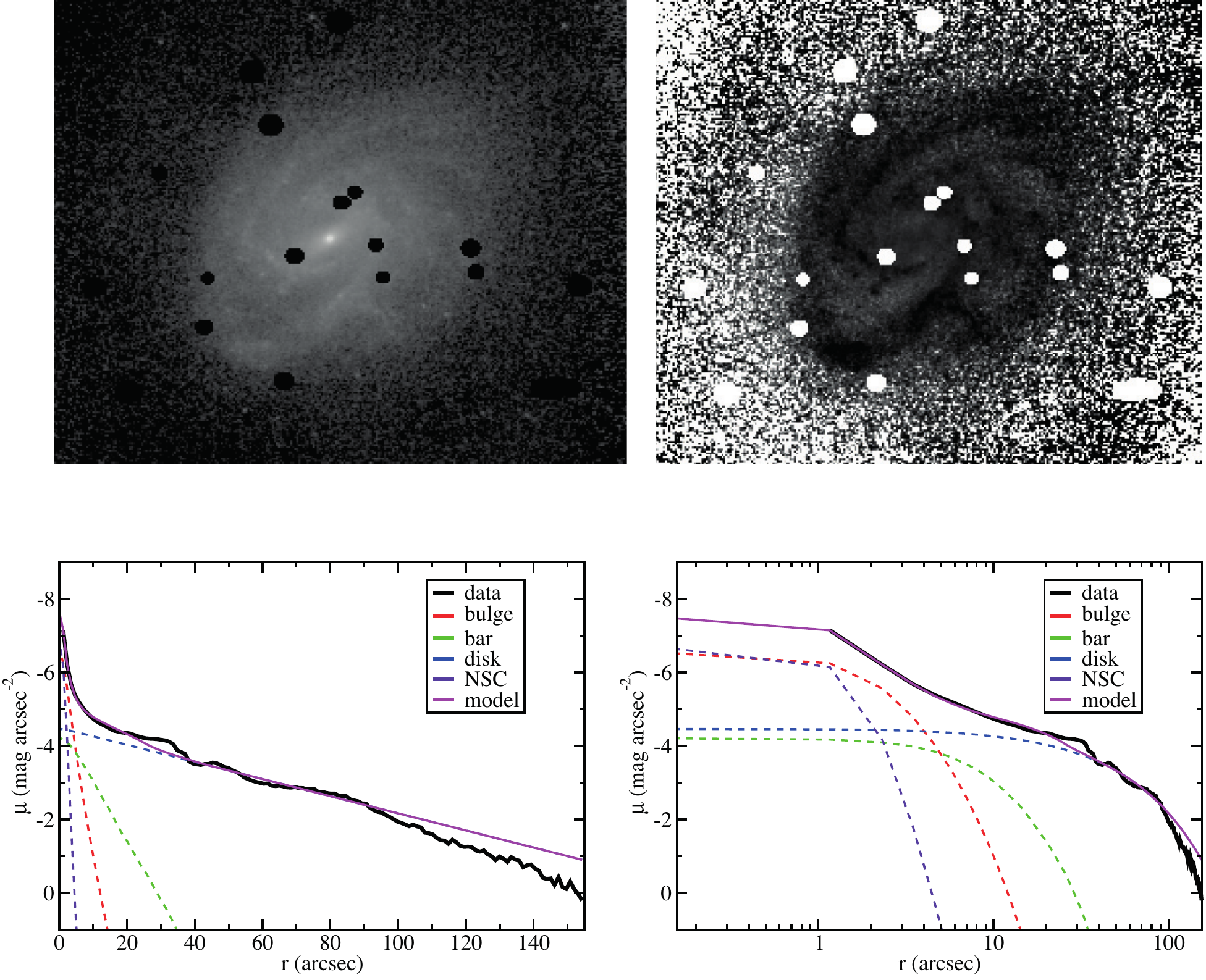}

\caption{An example of one of our 2D decompositions of NSC host galaxies -- in
this instance, the decomposition of the OSU BSGS $H$-band image of NGC 7418, using an
exponential disk, a S\'ersic bulge ($n = 1.5$), a bar, and a point source
for the NSC. \textbf{Upper left:} Original $H$-band image, with
masking of bright stars (logarithmic brightness scaling). \textbf{Upper right:}
Residual image after subtracting best-fitting model image. \textbf{Lower left:}
Major-axis profile (black) along with the components of the model and their
sum (purple). \textbf{Lower right:} Same, but plotted with logarithmic major-axis
scaling.}\label{fig:decomp2d}

\end{figure*}

Full details of the individual decompositions will be published elsewhere (Erwin
\& Gadotti 2012a, in prep.). An example of one of the 2D decompositions is given in
Figure~\ref{fig:decomp2d} for the galaxy NGC 7418, where we fit an $H$-band
image from the Ohio State University Bright Spiral Galaxy Survey \citep[OSU
BSGS]{osu} using an exponential disk (95.6\% of the light), a S\'ersic bulge
(1.6\% of the light), a bar (2.4\% of the light), and a point source for the NSC
(0.5\% of the light).\footnote{In this particular galaxy, the disk appears to be
truncated, but this has little effect on the decomposition; including a broken-exponential profile for the disk changes the $B/T$
ratio from 0.016 to 0.017.}
This illustrates the importance of including a separate
bar component in the decomposition when the galaxy is barred: the bar has almost
twice the luminosity of the bulge, and without it, the bulge luminosity (and
stellar mass) would certainly be overestimated.  In fact, a 1-D decomposition
for this galaxy gives a $B/T$ value almost twice as large (0.030); similar
results were found for four other barred galaxies in the sample, with mean $B/T$
values a factor of 2.1 times larger when the bar was omitted \citep[see
also][]{gadotti08}.

\section{Comparing Black Holes and Nuclear Star Clusters}\label{sec:comparison}

Although black-hole--bulge correlations are sometimes described as correlations
between the black hole mass and the host \textit{galaxy} mass (or luminosity as
a proxy for mass), this is really only true for elliptical galaxies, where the
entire galaxy is the ``bulge''. \citet{kormendy-gebhardt01} explicitly compared
$B$-band total and bulge luminosities for SMBH hosts and showed that the latter
provided a much better correlation.   Most recently, \citet{kormendy11} showed
for a larger, updated sample that SMBH masses in disk galaxies correlated much
better with (classical) bulge $K$-band luminosity than with the luminosity of
the disk component; this naturally suggests that total-galaxy
luminosity is unlikely to correlate well with SMBH mass when the galaxy is
disk-dominated.

\begin{figure*}
\centering
\includegraphics[width=19cm]{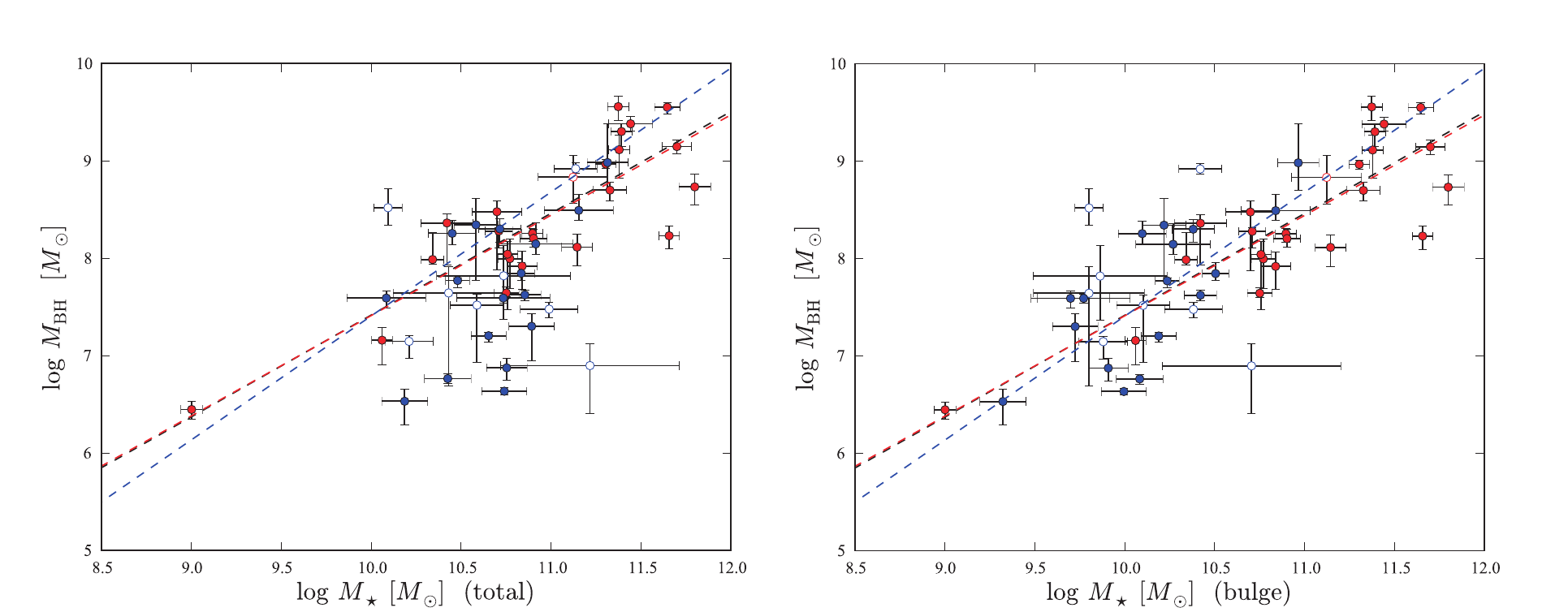}

  \caption{Left: SMBH mass (red = elliptical galaxies, blue = disk galaxies)
  versus total galaxy stellar mass. Right: SMBH mass versus bulge stellar mass.
  (Data and sources in Table~\ref{tab:bh}.)  
  The diagonal dashed lines are the best fits to the \mbh-\mbulge{} relation for 
  the whole sample (black), for the elliptical galaxies (red), and for bulges
  of the disk galaxies (blue). Open symbols are galaxies without precise
  distances, which are not used in the fits. It is clear that the SMBH masses
  of S0 and spiral galaxies (blue) correlate better with the \textit{bulge} 
  stellar mass than with total galaxy mass.}\label{fig:smbh}

\end{figure*}

In Figure~\ref{fig:smbh}, we compare SMBH masses with total galaxy stellar mass
(left panel) and with bulge stellar mass (right panel), based on our careful
bulge/disk/bar decompositions (see Table~\ref{tab:bh}).  Error bars include the
effects of uncertainties in the distance and in the $M/L$ and $B/T$ ratios. As
expected, the correlation between SMBH mass and \textit{bulge} mass is much
stronger than any correlation with total galaxy mass: the Spearman correlation
coefficients are $r_{S} = 0.71$ for the \mbh-\mbulge{} relation, versus 0.29 for
the \mbh-\mtot{} relation, with the latter correlation not being statistically
significant.

We also plot linear fits of $\log M_{\rm BH}$ as a function of $\log M_{\star,
\, {\rm bul}}$; these fits are made using galaxies with well-determined
distances (filled points) to minimize distance-based uncertainties, using the
Bayesian-based approach of \citet{dagostini05}, which explicitly incorporates
errors in both variables and intrinsic scatter in the black hole mass \citep[see
also][]{guidorzi06,gnedin07}.  By ``well-determined distances'' we mean those
determined using direct methods such as surface-brightness fluctuations and
Cepheid stars, or redshift-based distances where $z > 0.01$ (to avoid large
relative uncertainties due to peculiar motions.) The best-fitting relation for
the whole sample (black line; the fit to just the elliptical galaxies, shown by
the red line, is almost identical) is: 
\begin{equation}\label{eq:smbh-all} \log
\mbh \; = \; 8.46 \pm 0.08 \,+\, (1.04 \pm 0.12) \, \log(\mbulge/10^{11} \Msun),
\end{equation} 
with intrinsic scatter in SMBH mass of $0.39 \pm 0.05$ dex;  the
best-fitting relation for the bulges of disk galaxies only (blue line) is
\begin{equation} \log \mbh \; = \; 8.68 \pm 0.20 \,+\, (1.27 \pm 0.26) \,
\log(\mbulge/10^{11}  \Msun), 
\end{equation} 
with instrinsic scatter $= 0.41 \pm 0.07$. (The errors are based on bootstrap resampling.)

\begin{figure*}
\centering
\includegraphics[width=19cm]{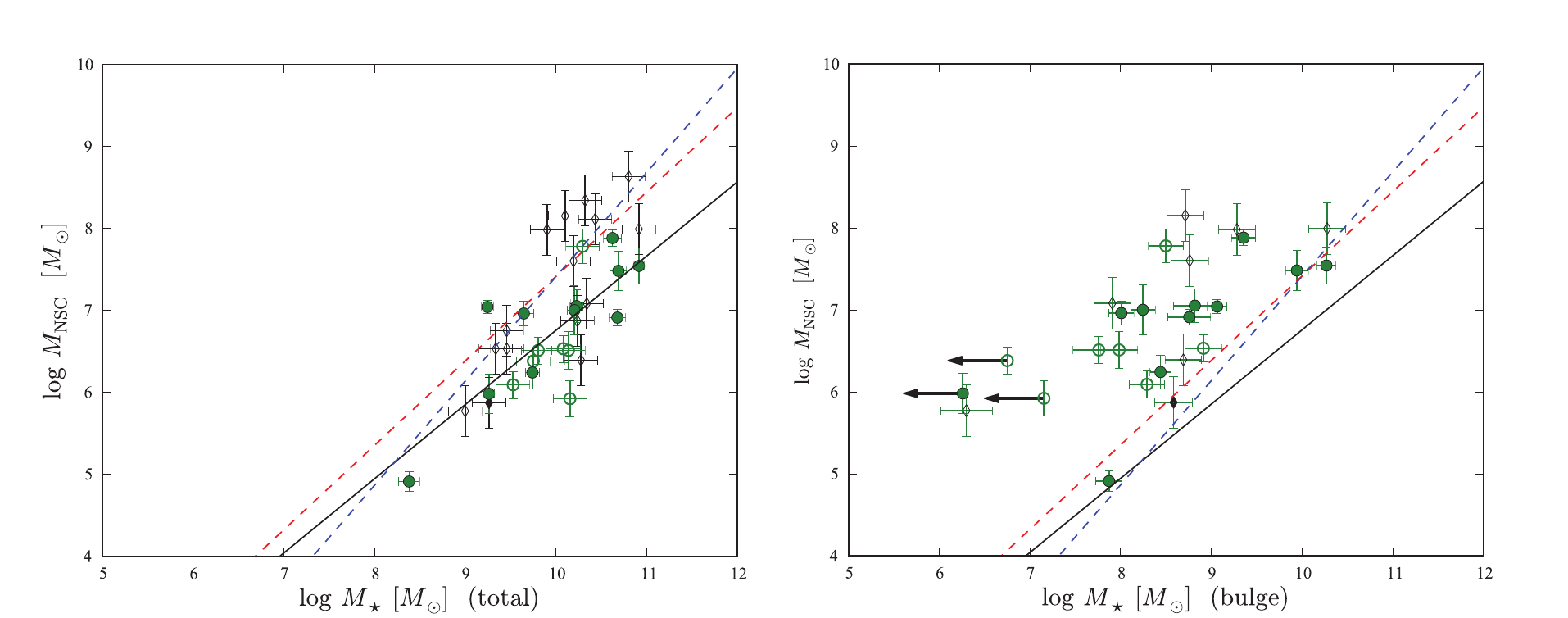}
  
\caption{As for Figure~\ref{fig:smbh}, but now plotting NSC mass versus total
stellar mass (left) and bulge stellar mass (right).  (Data and sources in Table~\ref{tab:nsc}.) 
Green circles are galaxies
with dynamical mass estimates for their NSCs; black diamonds are the
spectroscopically estimated masses of \citet{rossa06} (bulge mass estimates are not
complete for these galaxies).  Filled symbols indicate galaxies with direct
distance measurements (e.g., from Cepheid stars). Arrows show nominal upper
limits for three \textit{bulgeless} spirals (assuming $B/T \leq 0.001$). The
diagonal black line is a fit of NSC mass to total stellar mass for the
dynamical-mass sample (green circles); for comparison, the diagonal dashed red
and blue lines are the \mbh-\mbulge{} fits for ellipticals (red) and disk galaxies (blue)
from Figure~\ref{fig:smbh}. The situation is now the reverse of that for SMBHs:
NSC masses clearly correlate better with \textit{total} galaxy mass than they do
with bulge mass.}\label{fig:nsc-stellar-mass}
  
\end{figure*}

We apply exactly the same methodology to NSC-host galaxies in
Figure~\ref{fig:nsc-stellar-mass}, plotting NSC mass versus total galaxy stellar
mass in the left panel and versus bulge stellar mass in the right panel. Since
several of the NSC host galaxies are genuinely \textit{bulgeless} systems
(without even a distinct ``pseudobulge''), we plot their bulge masses as upper
limits ($B/T < 0.001 \mtot$). As the figure shows, NSC mass clearly
correlates better with \textit{total} stellar mass than it does with bulge mass.
(The respective correlation coefficients are $r_{S} = 0.76$ versus 0.38;
the bulge-mass correlation is not statistically significant.)  Fitting NSC
mass versus total stellar mass, using the same methodology as for the SMBH
fits, gives the following relation:
\begin{equation}
  \log \mnsc \; = \; 7.65 \pm 0.23 \,+\, (0.90 \pm 0.21) \, \log(\mtot/10^{11}  \Msun),
\end{equation}
with intrinsic scatter $= 0.43 \pm 0.10$ dex.  Note that the slope is formally
indistinguishable from unity: i.e., the $\mnsc/\mtot$ ratio does not appear to
depend on \mtot{} itself. 

It is important to note that the difference in correlation coefficients actually
\textit{underestimates} the true difference between the two relations, because
the \mnsc-\mbulge{} correlation was computed assuming that bulgeless spirals
still have nominal bulges (using $B/T = 0.001$).  In the combined sample of dynamical
and spectroscopic NSC masses, we can identify at least three galaxies which have
no detectable bulge.  In two of these (NGC 1493 and NGC 2139), our 2D
decomposition assigned stellar light to a bar in addition to a pure exponential
disk; in 1D decompositions (or simple bulge+disk 2D decompositions), light from
the bar might be (wrongly, we would argue) interpreted as ``bulge'' light.  For
the other galaxy (NGC 300), however, there is no ambiguity: this is an unbarred
spiral galaxy with a surface brightness profile consisting of \textit{only} an
exponential disk and the NSC \citep[see, e.g., Figure~8 of][]{bland-hawthorn05}.

The existence of nuclear star clusters in genuinely bulgeless spirals is simply
an additional, direct confirmation of our basic conclusion: nuclear star cluster
masses scale with the total stellar mass of their host galaxies, \textit{not}
with the bulge mass. This means that NSCs and SMBHs do not follow a common
host-galaxy correlation.



We have also investigated whether other galaxy parameters might
correlate with NSC mass, or even with residuals from the \mnsc-\mtot{}
relation. In particular, we have compared NSC mass with rotation velocity and
with total \textit{baryonic} mass (stellar mass plus atomic gas from \hi{}
measurements). In both cases, correlations exist, but they are not as strong as
the correlation with total stellar mass.  No particular correlations with
residuals of the \mnsc-\mtot{} relation are seen.

\section{Trends with Hubble Type}

Closer inspection of the left-hand panel of Figure~\ref{fig:nsc-stellar-mass}
suggests that the spectroscopic masses (black diamonds) tend to be offset from
the NSC-$\mtot$ relation, in the sense that they have larger NSC masses for
the same total stellar mass. This could, in principle, be evidence of a
systematic overestimation of NSC masses in the spectroscopic sample, but of the
four galaxies in common between \citet{walcher05} and \citet{rossa06} only one has
a (slightly) higher spectroscopic mass, while the other three have spectroscopic
masses slightly \textit{lower} than the dynamical masses.  There is, however,
another difference to consider: the spectroscopic sample tends to have earlier
Hubble types.

\begin{figure}
\resizebox{\hsize}{!}{\includegraphics{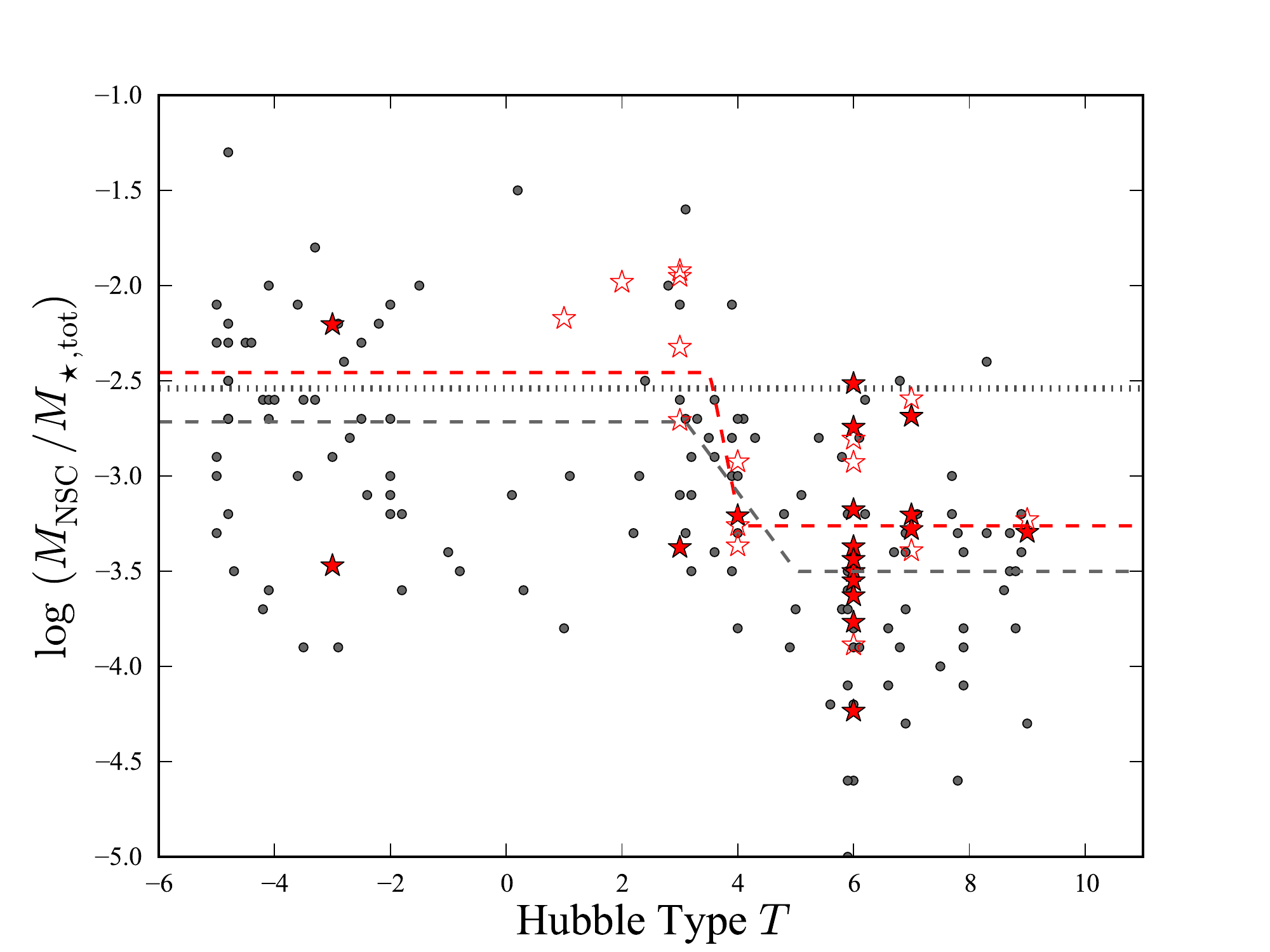}}
  
\caption{Relative masses of NSCs versus Hubble type of host galaxy, based on the
compilation of \citet{seth08}. Filled gray circles are NSC masses estimated from
broadband colors or assumed $M/L$ ratios by \citet{seth08}; red stars indicate
NSC masses from spectroscopic (hollow) or dynamical (filled) measurements (see
Table~\ref{tab:nsc} for references). Also shown are simple fits to the
dynamical+spectroscopic masses (dashed red line) and to the entire sample
(dashed gray line), along with the mean mass ratio of SMBHs relative to
their host \textit{bulges} (dotted gray line).}\label{fig:nsc-vs-hubble}
  
\end{figure}

This brings us to something which \citet{seth08} pointed out several years ago,
using a larger dataset of NSCs and host galaxies, with NSC masses based (mostly)
on colors or assumed $M/L$ ratios. They noted that NSCs in late-type spirals
tended to have lower relative masses ($\mnsc/\mtot$) than early-type spirals and
ellipticals.\footnote{\citet{rossa06} pointed out a similar trend in
\textit{absolute} NSC mass for their smaller sample of NSCs in early- and late-type
spirals.} Figure~\ref{fig:nsc-vs-hubble} makes this explicit by plotting
$\mnsc/\mtot$ versus Hubble type for the galaxies in Seth et al.'s compilation,
plus seven galaxies from our updated dynamical-mass sample which were not in
their sample. We have also added galaxy stellar-mass estimates for 16 galaxies
that did not have masses in Seth et al., using total $K$-band magnitudes from
2MASS and either $B - V$ colors from HyperLeda or measured $g - r$ colors from
SDSS images to derive the $K$-band $M/L$ via \citet{bell03}.

What is curious about Figure~\ref{fig:nsc-vs-hubble} is not just that the
$\mnsc/\mtot$ ratio depends on Hubble type, but that it actually appears do so
in a \textit{bimodal} fashion: Hubble types Sb and earlier have relatively large
NSC masses, while Sc and later-type galaxies have significantly smaller relative
NSC masses.  Plotted on top of the figure are simple fits of a function where
the the $\mnsc/\mtot$ ratio can take two constant values, one for Hubble types
$T < T_{1}$ and the other for $T > T_{2}$, with a simple linear transition
between $T_{1}$ and $T_{2}$. Fits to just the dynamical + spectroscopic masses
(red dashed line) and to the entire sample (gray dashed line) are similar,
indicating that Sb and earlier Hubble types form one class, with $<\mnsc/\mtot>
\sim 0.002$, and Sc and later types form a different group, with $<\mnsc/\mtot>$
almost an order of magnitude smaller ($\sim 0.0003$).  The corresponding
best-fit values of ($T_{1}$,$T_{2}$) are (3.51,4.05) for the dynamical +
spectroscopic masses and (3.10,5.01) for the complete sample. As a crude check
on whether this split is statistically significant, we performed
Kolmogorov-Smirnov tests on the values of $\mnsc/\mtot$ for galaxies with $T
\leq 3$ and galaxies with $T \geq 5$.  The K-S test gives a probability $P_{KS}
= 0.0038$ for the two sets of ratios coming from the same parent population if
we use only the dynamical + spectroscopic masses, or $P_{KS} = 3.1 \times
10^{-10}$ if we use the entire set of NSC masses.

Do NSCs in late-type spirals differ from those in early-type spirals, S0s, and
ellipticals in any sense other than average mass?  The available evidence is
ambiguous.  \citet{boker08} notes that NSC \textit{sizes} appear to be
independent of Hubble type. On the other hand, \citet{rossa06} compared stellar
populations of NSCs in early- and late-type spirals using fits to their
spectroscopy and noted that the NSCs in late-type spirals did tend to have
younger stellar populations and (slightly) lower metallicities. (They also
argued against any observational effects that might produce systematic
overestimates of NSC mass in early spirals.) This does at least suggest that
different star-formation histories may lie behind the mass differences in NSCs.

We also plot the $\mbh/\mbulge$ ratio for SMBH host galaxies (thick gray dotted
line in Figure~\ref{fig:nsc-vs-hubble}, based on Equation~\ref{eq:smbh-all}).
What this indicates is that the NSC--host galaxy relationship for Sb and earlier
types \textit{is} consistent with the SMBH relation, \textit{if all of the
galaxy mass is in the bulge}. Since most of the galaxies used for the original
CMO studies \citep{wehner06,ferrarese06} were dwarf and giant ellipticals (or S0
galaxies with high $B/T$ ratios), it is easy to see why the ``NSC = SMBH''
connection could be made. But this is clearly true only for very bulge-dominated
systems.

%
%

\begin{acknowledgements} 

We would like to thank Eva Noyola and Anil Seth for useful and interesting
conversations, along with comments from two referees which improved the
manuscript.  We would also like to thank the organizers of the 2008 ``Nuclear
Star Clusters across the Hubble Sequence'' workshop in Heidelberg and the 2010
``ESO Workshop on Central Massive Objects'' in Garching for helping motivate and
inspire this research. This work was supported by Priority Programme 1177
(``Witnesses of Cosmic History:  Formation and evolution of black holes,
galaxies and their environment'') of the Deutsche Forschungsgemeinschaft.

This work made use of data from the Ohio State University Bright Spiral
Galaxy Survey, which was funded by grants AST-9217716 and AST-9617006
from the United States National Science Foundation, with additional
support from Ohio State University. It is also based on observations
made with the NASA/ESA Hubble Space Telescope, obtained from the data
archive at the Space Telescope Science Institute.  STScI is operated by
the Association of Universities for Research in Astronomy, Inc.  under
NASA contract NAS 5-26555.

Funding for the creation and distribution of the SDSS Archive has been
provided by the Alfred P. Sloan Foundation, the Participating
Institutions, the National Aeronautics and Space Administration, the
National Science Foundation, the U.S. Department of Energy, the Japanese
Monbukagakusho, and the Max Planck Society.  The SDSS Web site is
http://www.sdss.org/.

The SDSS is managed by the Astrophysical Research Consortium (ARC) for
the Participating Institutions.  The Participating Institutions are The
University of Chicago, Fermilab, the Institute for Advanced Study, the
Japan Participation Group, The Johns Hopkins University, the Korean
Scientist Group, Los Alamos National Laboratory, the
Max-Planck-Institute for Astronomy (MPIA), the Max-Planck-Institute for
Astrophysics (MPA), New Mexico State University, University of
Pittsburgh, University of Portsmouth, Princeton University, the United
States Naval Observatory, and the University of Washington.

Finally, this research made use of the Lyon-Meudon Extragalactic
Database (LEDA; part of HyperLeda at http://leda.univ-lyon1.fr/), and
the NASA/IPAC Extragalactic Database (NED), which is operated by the
Jet Propulsion Laboratory, California Institute of Technology, under
contract with the National Aeronautics and Space Administration.

\end{acknowledgements}

\bibliographystyle{unsrtnat}

\appendix
\section{Data Tables}

In Tables~\ref{tab:bh} and \ref{tab:nsc}, we list the basic data parameters for SMBH and NSC
hosts. References for the NSC masses are in the captions for Table~\ref{tab:nsc}. For the
SMBH masses, the numbers in column 5 of Table~\ref{tab:bh} translate into the following
references: 1 = \citet{gillessen09}; 
2 = \citet{bender05}; 
3 = \citet{verolme02}; 
4 = \citet{krajnovic09}; 
5 = \citet{gebhardt03}; 
6 = \citet{bower01}; 
7 = \citet{lodato03}; 
8 = \citet{atkinson05}; 
9 = \citet{rusli11}; 
10 = \citet{nowak08}; 
11 = \citet{houghton06}; 
12 = \citet{sarzi01}; 
13 = \citet{devereux03}; 
14 = \citet{davies06}; 
15 = \citet{barth01}; 
16 = \citet{nowak10}; 
17 = \citet{kondratko08}; 
18 = \citet{gueltiken09a}; 
19 = \citet{defrancesco06}; 
20 = \citet{hicks08}; 
21 = \citet{miyoshi95}; 
22 = \citet{ferrarese96}; 
23 = \citet{cretton99}; 
24 = \citet{walsh10}; 
25 = \citet{macchetto97}; 
26 = \citet{nowak07}; 
27 = \citet{shen10}; 
28 = \citet{defrancesco08}; 
29 = \citet{neumayer07}; 
30 = \citet{capetti05};
31 = \citet{ferrarese99}; 
32 = \citet{vandermarel98};
33 = \citet{cappellari02}; 
34 = \citet{dalla-bonta09}.

\begin{table*}
\begin{minipage}{126mm}
    \caption{Galaxies with Well-Determined SMBH Masses}
    \label{tab:bh}
    \begin{tabular}{lrrrrrr}
\hline
Name & $T$ &  $D$  &   \mbh{} $(+,-)$        & Source     & Total \mstar  (err)     & Bulge \mstar (err) \\
     &     & (Mpc) &  $(\log_{10} \, \Msun)$ &            & $(\log_{10} \, \Msun)$  & $(\log_{10} \, \Msun)$    \\
\hline
Milky Way  &  4 & 0.01  & 6.63 (+0.03, $-0.04$)  & 1  & 10.74 (0.09)  & 10.00 (0.13)  \\
M31  &  3 & 0.77  & 8.15 (+0.22, $-0.10$)  & 2  & 10.92 (0.06)  & 10.27 (0.21)  \\
M32  &  -5 & 0.79  & 6.45 (+0.08, $-0.10$)  & 3  & 9.00 (0.06)  & 9.00 (0.06)  \\
NGC524  &  -1 & 23.3  & 8.92 (+0.04, $-0.02$)  & 4  & 11.14 (0.09)  & 10.42 (0.12)  \\
NGC821  &  -5 & 23.4  & 7.92 (+0.15, $-0.23$)  & 5  & 10.84 (0.08)  & 10.84 (0.08)  \\
NGC1023  &  -1 & 11.1  & 7.62 (+0.04, $-0.04$)  & 6  & 10.85 (0.08)  & 10.42 (0.09)  \\
NGC1068  &  3 & 14.3  & 6.90 (+0.14, $-0.21$)  & 7  & 11.22 (0.49)  & 10.71 (0.50)  \\
NGC1300  &  4 & 18.9  & 7.82 (+0.29, $-0.29$)  & 8  & 10.74 (0.37)  & 9.86 (0.37)  \\
NGC1316  &  -5 & 21.3  & 8.23 (+0.10, $-0.13$)  & 9  & 11.66 (0.06)  & 11.66 (0.06)  \\
NGC1399  &  -5 & 21.1  & 9.11 (+0.15, $-0.29$)  & 10  & 11.38 (0.06)  & 11.38 (0.06)  \\
NGC2549  &  -2 & 12.3  & 7.15 (+0.02, $-0.16$)  & 4  & 10.21 (0.12)  & 9.88 (0.13)  \\
NGC2748  &  4 & 23.1  & 7.64 (+0.25, $-0.74$)  & 8  & 10.43 (0.30)  & 9.80 (0.31)  \\
NGC2787  &  -1 & 7.28  & 7.59 (+0.04, $-0.06$)  & 11  & 10.08 (0.15)  & 9.70 (0.22)  \\
NGC3031  &  1 & 3.63  & 7.85 (+0.11, $-0.07$)  & 12  & 10.83 (0.06)  & 10.51 (0.07)  \\
NGC3227  &  1 & 22.9  & 7.30 (+0.13, $-0.35$)  & 13  & 10.89 (0.11)  & 9.72 (0.13)  \\
NGC3245  &  -1 & 20.3  & 8.30 (+0.10, $-0.12$)  & 14  & 10.72 (0.09)  & 10.38 (0.12)  \\
NGC3368  &  2 & 10.5  & 6.88 (+0.09, $-0.12$)  & 15  & 10.75 (0.09)  & 9.91 (0.11)  \\
NGC3377  &  -5 & 10.9  & 7.99 (+0.28, $-0.05$)  & 5  & 10.34 (0.06)  & 10.34 (0.06)  \\
NGC3379  &  -5 & 10.3  & 8.00 (+0.20, $-0.31$)  & 5  & 10.77 (0.07)  & 10.77 (0.07)  \\
NGC3384  &  -1 & 11.3  & 7.20 (+0.03, $-0.05$)  & 5  & 10.65 (0.08)  & 10.19 (0.10)  \\
NGC3393  &  1 & 48.3  & 7.48 (+0.03, $-0.03$)  & 16  & 10.99 (0.15)  & 10.38 (0.16)  \\
NGC3489  &  -1 & 11.7  & 6.76 (+0.04, $-0.04$)  & 15  & 10.43 (0.08)  & 10.08 (0.13)  \\
NGC3585  &  -3 & 19.5  & 8.49 (+0.16, $-0.09$)  & 17  & 11.15 (0.09)  & 10.84 (0.19)  \\
NGC3607  &  -5 & 22.2  & 8.11 (+0.13, $-0.19$)  & 17  & 11.15 (0.08)  & 11.15 (0.08)  \\
NGC3608  &  -5 & 22.3  & 8.28 (+0.02, $-0.16$)  & 5  & 10.71 (0.08)  & 10.71 (0.08)  \\
NGC3998  &  -2 & 13.7  & 8.34 (+0.27, $-0.56$)  & 18  & 10.58 (0.09)  & 10.22 (0.12)  \\
NGC4026  &  -3 & 13.2  & 8.26 (+0.12, $-0.09$)  & 17  & 10.45 (0.12)  & 10.10 (0.13)  \\
NGC4151  &  2 & 14.5  & 7.52 (+0.10, $-0.56$)  & 19  & 10.59 (0.13)  & 10.10 (0.15)  \\
NGC4258  &  4 & 7.18  & 7.59 (+0.04, $-0.04$)  & 20  & 10.73 (0.08)  & 9.77 (0.26)  \\
NGC4261  &  -5 & 30.8  & 8.70 (+0.08, $-0.10$)  & 21  & 11.33 (0.09)  & 11.33 (0.09)  \\
NGC4291  &  -5 & 25.5  & 8.48 (+0.10, $-0.57$)  & 5  & 10.70 (0.14)  & 10.70 (0.14)  \\
NGC4342  &  -1 & 16.7  & 8.52 (+0.20, $-0.18$)  & 22  & 10.09 (0.07)  & 9.80 (0.08)  \\
NGC4374  &  -5 & 18.5  & 8.97 (+0.04, $-0.04$)  & 23  & 11.30 (0.06)  & 11.30 (0.06)  \\
NGC4473  &  -5 & 15.3  & 8.04 (+0.13, $-0.56$)  & 5  & 10.76 (0.06)  & 10.76 (0.06)  \\
NGC4486  &  -5 & 16.7  & 9.56 (+0.11, $-0.14$)  & 24  & 11.37 (0.06)  & 11.37 (0.06)  \\
NGC4486A  &  -5 & 18.4  & 7.16 (+0.13, $-0.25$)  & 25  & 10.06 (0.06)  & 10.06 (0.06)  \\
NGC4564  &  -3 & 15.9  & 7.77 (+0.02, $-0.07$)  & 5  & 10.48 (0.06)  & 10.24 (0.07)  \\
NGC4649  &  -5 & 16.4  & 9.30 (+0.08, $-0.15$)  & 26  & 11.39 (0.06)  & 11.39 (0.06)  \\
NGC4697  &  -5 & 12.5  & 8.26 (+0.05, $-0.09$)  & 5  & 10.90 (0.06)  & 10.90 (0.06)  \\
NGC5077  &  -5 & 37.5  & 8.83 (+0.21, $-0.23$)  & 27  & 11.12 (0.19)  & 11.12 (0.19)  \\
NGC5128  &  -5 & 3.42  & 7.64 (+0.06, $-0.03$)  & 28  & 10.75 (0.07)  & 10.75 (0.07)  \\
NGC5252  &  -2 & 92.9  & 8.98 (+0.40, $-0.27$)  & 29  & 11.31 (0.09)  & 10.97 (0.11)  \\
NGC5576  &  -5 & 24.8  & 8.20 (+0.09, $-0.08$)  & 17  & 10.90 (0.08)  & 10.90 (0.08)  \\
NGC5845  &  -5 & 25.2  & 8.36 (+0.07, $-0.41$)  & 5  & 10.42 (0.15)  & 10.42 (0.15)  \\
NGC6251  &  -5 & 95.9  & 8.73 (+0.12, $-0.18$)  & 30  & 11.80 (0.09)  & 11.80 (0.09)  \\
NGC7052  &  -5 & 67.9  & 8.58 (+0.23, $-0.22$)  & 31  & 11.49 (0.12)  & 11.49 (0.12)  \\
NGC7457  &  -1 & 12.9  & 6.53 (+0.12, $-0.23$)  & 5  & 10.19 (0.10)  & 9.32 (0.13)  \\
IC1459  &  -5 & 28.4  & 9.38 (+0.05, $-0.04$)  & 32  & 11.44 (0.12)  & 11.44 (0.12)  \\
IC4296  &  -5 & 53.2  & 9.15 (+0.06, $-0.07$)  & 33  & 11.70 (0.08)  & 11.70 (0.08)  \\
A1836-BCG  &  -5 & 155.6  & 9.55 (+0.05, $-0.06$)  & 33  & 11.65 (0.07)  & 11.65 (0.07)  \\
\hline
\end{tabular}

(1) Galaxy name.  (2) Hubble type $T$ from RC3. (3) Adopted distance in Mpc. (4) 
Logarithm of SMBH mass and uncertainties;
masses have been rescaled using the distances column 2, if necessary.  Uncertainties
are 1-$\sigma$ values. (5) Source of SMBH measurement.  (6) Logarithm of total
galaxy stellar mass and uncertainty (see text for details).  (7) Logarithm of bulge stellar
mass and uncertainty, based on 2D decompositions in Erwin \& Gadotti (2012b, in prep).
   
\end{minipage}
\end{table*}

\begin{table*}
\begin{minipage}{126mm}
    \caption{Galaxies with Well-Determined NSC Masses}
    \label{tab:nsc}
    \begin{tabular}{lrrrrrrrr}
\hline
Name & $T$ &  $D$  &   \mnsc{} (err)          & Type    & Source  & Total \mstar (err)   & Bulge \mstar (err) \\
     &     & (Mpc) &  $(\log_{10} \, \Msun)$     &         &         & $(\log_{10} \, \Msun)$  & $(\log_{10} \, \Msun)$ \\
\hline
Milky Way  &  4 & 0.01  & 7.48 (0.09)  & D  & 1  & 10.74 (0.09)  & 10.00 (0.13)  \\
M31  &  3 & 0.77  & 7.54 (0.06)  & D  & 2  & 10.92 (0.06)  & 10.27 (0.11)  \\
M33  &  6 & 0.81  & 6.24 (0.08)  & D  & 3  & 9.74 (0.08)  & 8.44 (0.12)  \\
IC342  &  6 & 3.37  & 7.05 (0.07)  & D  & 4  & 10.23 (0.07)  & 8.82 (0.23)  \\
NGC300  &  7 & 2.02  & 5.98 (0.06)  & D  & 5  & 9.26 (0.06)  & $<$ 6.26  \\
NGC404  &  -3 & 3.18  & 7.04 (0.06)  & D  & 6  & 9.24 (0.06)  & 9.06 (0.11)  \\
NGC428  &  9 & 15.5  & 6.51 (0.18)  & D  & 5  & 9.81 (0.18)  & 7.76 (0.29)  \\
NGC1042  &  6 & 17.5  & 6.51 (0.18)  & D  & 5  & 10.14 (0.18)  & 7.98 (0.21)  \\
NGC1493  &  6 & 11.0  & 6.38 (0.19)  & D  & 5  & 9.75 (0.19)  & $<$ 6.75  \\
NGC1705  &  -3 & 5.11  & 4.91 (0.12)  & D  & 7  & 8.38 (0.12)  & 7.87 (0.15)  \\
NGC2139  &  6 & 22.9  & 5.92 (0.18)  & D  & 5  & 10.15 (0.18)  & $<$ 7.15  \\
NGC3423  &  6 & 14.4  & 6.53 (0.19)  & D  & 5  & 10.08 (0.19)  & 8.91 (0.21)  \\
NGC3621  &  7 & 6.64  & 7.00 (0.08)  & D  & 8  & 10.20 (0.08)  & 8.25 (0.14)  \\
NGC5457  &  6 & 7.05  & 6.91 (0.09)  & D  & 9  & 10.68 (0.09)  & 8.76 (0.24)  \\
NGC6946  &  6 & 5.89  & 7.88 (0.10)  & D  & 9  & 10.62 (0.10)  & 9.36 (0.13)  \\
NGC7418  &  6 & 17.8  & 7.78 (0.18)  & D  & 5  & 10.29 (0.18)  & 8.50 (0.19)  \\
NGC7424  &  6 & 10.5  & 6.09 (0.18)  & D  & 5  & 9.53 (0.18)  & 8.29 (0.19)  \\
NGC7793  &  7 & 3.91  & 6.96 (0.11)  & D  & 5  & 9.65 (0.11)  & 8.01 (0.14)  \\
NGC1325  &  4 & 19.6  & 7.08 (0.18)  & S  & 10  & 10.34 (0.18)  & 7.91 (0.20)  \\
NGC1385  &  6 & 18.1  & 6.39 (0.18)  & S  & 10  & 10.28 (0.18)  & 8.69 (0.20)  \\
NGC2552  &  9 & 9.68  & 5.77 (0.18)  & S  & 10  & 9.00 (0.18)  & 6.30 (0.28)  \\
NGC3177  &  3 & 19.6  & 8.15 (0.18)  & S  & 10  & 10.10 (0.18)  & 8.71 (0.20)  \\
NGC3277  &  2 & 21.4  & 8.34 (0.18)  & S  & 10  & 10.32 (0.18)  & \ldots{}  \\
NGC3455  &  3 & 16.4  & 6.75 (0.18)  & S  & 10  & 9.46 (0.18)  & \ldots{}  \\
NGC4030  &  4 & 20.5  & 7.99 (0.18)  & S  & 10  & 10.92 (0.18)  & 10.28 (0.20)  \\
NGC4411B  &  6 & 18.6  & 6.53 (0.19)  & S  & 10  & 9.46 (0.19)  & \ldots{}  \\
NGC4701  &  6 & 10.8  & 6.53 (0.18)  & S  & 10  & 9.34 (0.18)  & \ldots{}  \\
NGC4775  &  7 & 21.9  & 7.60 (0.19)  & S  & 10  & 10.19 (0.19)  & 8.76 (0.20)  \\
NGC5377  &  1 & 28.4  & 8.63 (0.18)  & S  & 10  & 10.80 (0.18)  & \ldots{}  \\
NGC5585  &  7 & 8.71  & 5.87 (0.19)  & S  & 10  & 9.26 (0.19)  & 8.58 (0.21)  \\
NGC5806  &  3 & 20.0  & 8.11 (0.18)  & S  & 10  & 10.43 (0.18)  & \ldots{}  \\
NGC7421  &  4 & 23.1  & 6.87 (0.18)  & S  & 10  & 10.24 (0.18)  & \ldots{}  \\
NGC7690  &  3 & 17.7  & 7.98 (0.18)  & S  & 10  & 9.90 (0.18)  & 9.28 (0.20)  \\
\hline
\end{tabular}

(1) Galaxy name.  (2) Hubble type $T$ from RC3. (3) Adopted distance in Mpc. (4) Logarithm of NSC mass and
uncertainty; masses have been rescaled using the distances column 2, if
necessary.  Errors are 1-$\sigma$ values. (5) Type of NSC mass measurement: ``D'' = dynamical,
``S'' = spectroscopic. (6) Source of NSC measurement: 1 = \citet{launhardt02}; 2 = \citet{kormendy-bender99};
3 = \citet{matthews99} + \citet{gebhardt01}; 4 = \citet{boker99};
5 = \citet{walcher05}; 6 = \citet{seth10}; 7 = \citet{ho96}; 8 = \citet{barth09}; 9 = \citet{kormendy10};
10 = \citet{rossa06}.  (7) Logarithm of total galaxy stellar mass and uncertainty (see text for details).  (8) Logarithm of
bulge stellar mass and uncertainty (or upper limit for bulgeless galaxies), based on decompositions in Erwin \& Gadotti (2012a, in
prep); galaxies currently missing proper decompositions are indicated by ``\ldots''.
   
\end{minipage}
\end{table*}

\end{document}